\documentclass{aa}
\usepackage{graphics}
\begin{document}

\title{Coronal abundances from high-resolution X-ray data:
\break The case of Algol}

\author{J.H.M.M. Schmitt\and J.-U. Ness}
\institute{
Hamburger Sternwarte, Universit\" at Hamburg, Gojenbergsweg 112,
D-21029 Hamburg, Germany\\
email: jschmitt@hs.uni-hamburg.de}

\offprints{J.H.M.M. Schmitt, \email{jschmitt@hs.uni-hamburg.de}}
\date{accepted: 19. September 2003}

\abstract{We discuss the determination of elemental abundances from
high resolution X-ray data. We emphasize the need for an accurate
determination of the underlying temperature structure and advocate
the use of a line ratio method which allows us to utilize, first, the
strongest lines observed in the X-ray spectra, and second, lines that
span a rather wide temperature range. We point out the need
to use continuous emission measure distributions and show via example
that modeling in terms of individual temperature components yields
errors of more than 50\%. We stress the need to derive differential
emission measure distributions based on physical assumptions and
considerations. We apply our methods to the {\it Chandra} LETGS
spectrum of Algol and show that nitrogen is considerably enhanced
compared to cosmic abundances by a factor of 2 while carbon is depleted 
by at least a factor of 25. Iron, silicon, and magnesium,
are all depleted compared to cosmic
abundances, while the noble gas neon has the relatively highest
abundance.
}

\maketitle

\section{Introduction}

The determination of elemental abundances in a variety of astrophysical
objects belongs to the most important tasks of observational astronomy and
the understanding of the evolution of chemical elements with cosmic
time is among the central themes of modern astrophysics. Elemental abundances
can be measured either from absorption spectra of stellar atmospheres or
from an analysis of the line emission spectrum of nebular emission. In both
cases the temperature structure of the emitting object must be known before
elemental abundances can be determined because for a given set of abundances
plasma temperature is (often) the most important factor controlling the
ionization equilibrium and hence the amount of a given type of material,
say O\,{\sc vii} or Fe\,{\sc xvii}, in an astrophysical object.

Reliable determinations of chemical abundances are carried out from high
resolution spectra. While elemental abundance determinations of stellar
photospheres can also be made from a set of suitably chosen filters,
abundances determined from high-resolution spectra are thought to be much
more accurate and far less model-dependent. This also and specifically
applies to X-ray data. The energy losses of hot thermal plasmas with
temperatures above $\approx$ 10$^6$\,K peak in the X-ray range and
therefore the chemical composition of such plasmas, which are encountered
in stellar coronae, in supernova remnants, and clusters of
galaxies, can in fact only be determined from X-ray data.

Continuum energy losses dominate the cooling of thermal plasmas above
$\approx$ 10$^7$\,K, while the thermal energy losses of plasma with
temperatures below $\approx$ 10$^7$\,K are dominated by a multitude of
emission lines. At those temperatures the strongest coolants are typically
(albeit not exclusively) the hydrogen- and helium-like ions of the most
abundant species, i.e., carbon, nitrogen, oxygen, neon, magnesium, silicon,
sulfur, calcium, argon, and iron. The hydrogen- and helium-like ions of the
elements heavier than sulfur are located below 6\,\AA, and are therefore
difficult to observe with high spectral resolution. Also, at temperatures
below $\approx$ 10$^7$\,K, the dominant stage of ionization for the heavier
atomic species is not yet advanced to helium- or hydrogen-like ions. For
example, for iron the most abundant stage of ionization at 10~MK is boron-like
Fe\,{\sc xxiii}, at a temperature of 3~MK neon-like Fe\,{\sc xvii} (cf.
Arnaud and Rothenflug \cite{ar85}), and consequently the energy losses from
iron are dominated by line emission from those ions.

The wealth of emission lines from trace elements in the X-ray range
demonstrates the potential of abundance determinations from such data.
Many calculations of the total energy output of a hot collisionally
ionized plasma have been made (cf. Raymond and Smith \cite{rs77}, Mewe et al.
\cite{mewe85}, Dere et al. \cite{dere97}), and the results of such
calculations have been used to interpret broad-band X-ray data as available
from the {\it Einstein Observatory} IPC or the ROSAT PSPC and HRI. Such plasma
codes have been used to infer an energy flux from a given count rate
measurement as well as to model the typically low-resolution pulse height
spectra of proportional, gas scintillation or CCD detectors. For the spectral
modeling individual spectral components are used. While the lowest resolution
data can be modeled with one or two temperature components with solar
abundances, higher resolution spectra require three or more temperature
components with typically non-solar abundances. X-ray detectors tend to work
very efficiently at energies of $\approx 1$\,keV, where both the effective
area of the instrument and the plasma emissivity from iron is at maximum.
Deviations from solar abundances in the X-ray spectra of stars were first
reported on the basis of ASCA CCD spectra of stars like Algol (cf. Antunes
et al. \cite{ant94}) and AR Lac (cf. White et al. \cite{whi94}); also the
even lower resolution PSPC spectra of some active stars were found to be
better fitted with metal-depleted rather than solar-abundance plasma models
(for example, Algol, cf. Ottmann and Schmitt \cite{ott96}, and CF~Tuc, cf.
Schmitt et al. \cite{schm96}). During large flares abundance changes were
inferred on the basis of spectral modeling of the time-dependent X-ray spectra.
Both in AB Dor (cf. G\"udel et al. \cite{gue2001}) as well as Algol (cf.
Favata and Schmitt \cite{fav99}) the iron abundance was found to increase
during the early rise phase of a flare, and then to decrease back no "normal"
sub-solar abundance values.

Abundance determinations of stellar coronae based on an analysis of
individual emission lines were first carried out with data from the
spectrometers on board the EUVE satellite. The emission line studies based
on EUVE data followed relatively closely the example of abundance
determinations of the solar corona. Stern et al. (\cite{stern95})
and Schmitt et al. (\cite{schm96}) found an anomalously low iron abundance
in the EUVE spectra of Algol and CF Tuc, based on an analysis of the
Fe\,{\sc xx}, Fe\,{\sc xxi}, and Fe\,{\sc xxii} emission lines in the XUV
range and the observed continuum values. Schmitt et al. (\cite{schm96})
coined the term metal abundance deficiency syndrome (MADS) for this
phenomenon, which is in contrast to the abundance pattern observed in the
solar corona, where elements with low first ionization potential (FIP) are
found to be enhanced. Drake et al. (\cite{drake96}) studied the presence
and absence of this so-called FIP-effect in a small number of nearby stars.
Using {\it Chandra} HETGS data Drake et al. (\cite{drake01}) study the
elemental abundance of the active binary HR\,1099 by means of a differential
emission
measure distribution computed from a Markov chain, while Brinkman et al.
(\cite{bri2001}) study the same source with the XMM-Newton reflection
grating spectrometer (RGS) assigning all of the emission measure to the
temperature corresponding to the peak of the line contribution function.
Other abundance studies based on {\it Chandra} or XMM-Newton include
Audard et al. (\cite{aud01}), who use a fit approach based on Chebyshev
polynomials and Huenemoerder et al. (\cite{hue2001}), who use a smoothed
positive emission measure distribution function.
G\"udel et al. (\cite{gue2001b}) use a fitting approach based on
individual temperature components to derive the elemental abundances in
YY~Gem using again data from the XMM-Newton RGS.

The purpose of this paper is to apply the ideas developed in solar and
stellar ultraviolet emission line studies to the now available broad band
and high spectral resolution X-ray data. The specific advantage of those data
coming from the recent generation of X-ray spectrometers on board {\it Chandra}
and XMM-Newton is that they cover the resonance lines of the
hydrogen- and helium-like ions of the elements carbon, nitrogen, oxygen,
neon, magnesium, and silicon. Resonance lines of the
hydrogen- and helium-like ions are among the strongest observable lines
and can be detected also in the weaker sources. Also, the atomic physics
required for an interpretation of those lines is - probably - simpler than
that required for the lines in more complex ions. We further follow quite
strictly the approach that the determination of the emission measure
distribution must occur in an abundance independent fashion, i.e., one
either uses only lines from a given element (in practice only iron can be
used) or one uses ratios of lines from the same element. The latter
approach has the enormous advantage that the strongest (rather than the
weaker) lines in an observed X-ray spectrum can be used for abundance
determinations, once the overall continuum emission level (or possibly that
of a well defined atomic species) has been fixed. 

The plan of our paper is therefore as follows: We first collect the
necessary formulae required for the calculation of line and continuum
fluxes from an isothermal plasma at temperature $T$ with specific emphasis on
the abundance dependence of these quantities. We introduce the concept
of the differential emission measure (DEM); the DEM distribution is
modeled by an approximation with Chebyshev polynomials on the one hand
and with the help of a Gaussian distribution of magnetic loops on the
other hand. The abundances computed in this fashion are juxtaposed to
those computed from a more conventional analysis with individual temperature
components.

\section{Abundance determination from optically thin plasma emission}

\subsection{Differential emission measure distribution}

\subsubsection{Line emission}

In this section we review the basic physics of coronal line formation
in as much as relevant for elemental abundance determinations.
The basic ideas of analysis have been summarized by Pottasch
(\cite{pott65}) in a solar context.
Consider the simplest case of a two-level atom in coronal equilibrium.
Coronal equilibrium implies an equilibrium between collisional excitation
from the lower level $l$ followed by radiative de-excitation from the excited
level $u$. The emitted photon leaves the system so that in essence
energy from the thermal pool has been converted into radiation. The
equilibrium condition then reads as follows:
\begin{equation}
n_{el} n_{ion,l} C_{lu} = n_{ion,u} A_{ul}\,, \label{l1}
\end{equation}
where $n_{el}, n_{ion,l}$, and $n_{ion,u}$ denote the number densities of
electrons and ions in lower and upper state, respectively, $C_{lu}$ is the
collisional rate coefficient and $A_{ul}$ the {\it Einstein}
for spontaneous radiative deexcitation. The total power emitted
in the transition $ul$ is given by
\begin{equation}
P_{ul} =  n_{ion,u} A_{ul} V_{corona}\,,\label{l2}
\end{equation}
where $V_{corona}$ denotes the total (isothermal) coronal volume. Let us
further assume that the considered transition is produced by a Z-times ionized
atom from atomic species $X$; in a low density environment almost all of these
ions will be in the ground state $l$, i.e., $n_{ion,l} = n_{X}^{Z+}$. We can
therefore express $n_{ion,l}$ in the following way:
\begin{equation}
n_{ion,l} = \frac {n_{X}^{Z+}} {n_X} \frac {n_X} {n_H}
\frac {n_H} {n_{el}} \ n_{el}\,, \label{l3}
\end{equation}
where $n_{X}$ denotes the total number density of all ions of atomic species
$X$ and $n_{H}$ the hydrogen/proton density. The abundance of species $X$
relative to hydrogen is denoted by $A_X$ and is given by
\begin{equation}
A_X = \frac {n_X} {n_H}\,, \label{l4}
\end{equation}
and in a more or less fully ionized cosmic abundance plasma we have
\begin{equation}
n_p = 0.85\ n_{el}\,.\label{l5}
\end{equation}
Note that for all plasmas with ``reasonable'' composition most of the
electrons come from hydrogen and helium, so that Eq.~\ref{l5} is virtually
independent of the assumed elemental abundances.
With these definitions the emitted power $P_{ul}$ can be written as
\begin{equation}
P_{ul} = 0.85\ n_{el}^2\ V_{corona}\ C_{lu} \frac {n_{X}^{Z+}} {n_X}\
A_X\,. \label{l6}
\end{equation}
The product $C_{lu} \frac {n_{X}^{Z+}}{n_X}$ depends only on temperature;
we can therefore write
\begin{equation}
P_{ul} = \Lambda _{ul}(T) EM A_X\,, \label{l7}
\end{equation}
where we have defined the volume emission measure $EM$ in the usual way
through
\begin{equation}
EM = n_{el}^2 V_{corona}\label{l8}
\end{equation}
and denote by $\Lambda _{ul}(T)$ the so-called line cooling function of the
transition $ul$. Eq.~\ref{l7} illustrates the fundamental dilemma of
all abundance determinations using optical thin emission in coronal
equilibrium: Cooling function, emission measure, and abundance all enter in
the same multiplicative fashion, and therefore one has to know both emission
measure and temperature in order to determine the abundance.
The temperature dependence of the collisional excitation coefficients
is typically of the form $C_{lu} \sim T^{-1/2} e^{-\chi/kT}$, with $\chi$
denoting the line's excitation potential, and similar functional dependencies
apply to the temperature dependence of the fractional ionization of
a given ion. Consequently, a given line is sensitive over
a relatively broad range of temperature with a width of typically 0.3 dex.
\\
So far we have assumed isothermality. Stellar coronae are very likely not
isothermal, and different coronal volume elements contributing to a given
observed line flux in the transition $u$ to $l$ have in general
different temperatures. Consider an infinitesimal
volume element $dV$ in a temperature range $dT$, and the differential power
$dP_{ul}$ emitted from this volume element. The total power $P_{ul}$ can
then be computed in the following way:
\begin{eqnarray}
\label{l9}
P_{ul} = \int dP_{ul} = A_X \int dV n_{el}^2 \Lambda _{ul}(T) =\\
  A_X \int dT \xi(T) \Lambda _{ul}(T)\,,&&\nonumber
\end{eqnarray}

where temperature $T$ was used to replace the integration variable $V$.
The quantity $\xi(T)$ is defined through
\begin{equation}
\xi(T) = n_{el}^2(T) \frac {dV} {dT}\label{l10}
\end{equation}
and is known as the so-called differential emission measure (DEM). Obviously
we must know the differential emission measure $\xi(T) $ in order to
determine the abundance $A_X$. Eq.~\ref{l9} shows, however, that the observed
line flux $ P_{ul}$ is an integral of the product of the line cooling function
and the differential emission measure distribution. If a number of
different lines with cooling functions $\Lambda _{ul}(T)$ with different
temperature sensitivity are available, one obtains an integral equation
with the differential emission measure distribution $\xi(T)$ as kernel. The
system is a Fredholm equation of the first kind. Such integral equations are
notoriously difficult to solve and, in particular, need not have unique solutions.
This has been known to mathematicians for a long time,
Craig and Brown (\cite{cb76}) were the first to point out the
resulting limitations for our ability to extract physical information
from spectral analysis in a cogent form.

\subsubsection{Continuum emission}

We now consider continuum radiation from a hot plasma. Continuum
emission comes from (free-free) bremsstrahlung, from free-bound radiation,
and two photon radiation; for temperatures
$\ge$ 5\,MK thermal bremsstrahlung is the dominant continuum energy loss
mechanism. All these continuum emission processes originate from interactions
of either protons or ions with free electrons very similar to the generation
of line emission. One thus expects that the dependence of the continuum
emission on electron density and temperature is of the same functional form
as for line emission. Indeed,
Mewe et al. (\cite{mewe86}) consider an isothermal plasma with electron number density
$n_e$ at some temperature $T$ and write the specific continuum emissivity as
\begin{equation}
\frac {dW} {dV dt d\lambda} = \frac {2^5 \pi e^6} {3 m c^2}
\sqrt{\frac {2 \pi} {3 k m T}} n_e^2 \frac {1} {\lambda ^2}
e^{-\frac {h c} {\lambda k T}}
g_{c}(\lambda, T); \label{c1}
\end{equation}
here $h$ and $k$ denote Planck's and Boltzmann's constants, respectively.
The quantity $g_{c}(\lambda, T)$ is the velocity-averaged
Gaunt factor. It can be written as a sum of three components through
\begin{equation}
g_{c} = g_{ff} + g_{fb} + g_{2\gamma}, \label{c1a}
\end{equation}
where $g_{ff}$, $g_{fb}$, and $g_{2\gamma}$ represent the contributions
from free-free, free-bound, and two photon radiation.

The different constituents of the continuum emissivity do, however,
have different
dependences on elemental abundances. The bremsstrahlung component
comes predominantly from electron-proton collisions with most electrons
due to fully ionized hydrogen and helium atoms. Under ``reasonable''
abundances the number of electrons from all heavier elements will be
very small. In contrast, the other components do depend on trace
element abundances
since the free-bound radiation depends on the recombination frequency and
the two-photon radiation on the number density of meta-stable He-like
states, both of which are abundance dependent. Fortunately, for a sufficiently
hot plasma electron bremsstrahlung is dominant so at least in first order the
continuum energy loss is abundance independent. The specific power
$dP_{\lambda}$ emitted from a volume element $dV$ is thus given by
\begin{equation}
dP_{\lambda} = \frac {2^5 \pi e^6} {3 m c^2} \sqrt{\frac {2 \pi} {3 k m T}}
n_e^2 e^{-\frac {h c} {\lambda k T}} \frac{1}{\lambda^2} g_{c}(\lambda, T) dV. \label{c3}
\end{equation}
The specific power emitted from a continuous emission measure distribution
described by a distribution function $\xi(T)$ is then given by
\begin{equation}
P_{\lambda} = \frac {2^5 \pi e^6} {3 m c^2}
\int \sqrt{\frac {2 \pi} {3 k m T}} e^{-\frac {h c} {\lambda k T}} \frac{1}
{\lambda^2} g_{c}(\lambda, T)
\xi (T) dT. \label{c4}
\end{equation}

Thus from a mathematical point of view, the continuum emissivity is again
given by an integral of the differential emission measure distribution
function $\xi(T)$ with yet another kernel. In contrast to line emission,
however, the continuum emission does, first, not depend sensitively on the
plasma's metal abundance, and second, its temperature dependence is very different.
At high temperatures, where bremsstrahlung dominates, the continuum
emission is virtually independent of metal abundance since almost all electrons
in a plasma come from the lightest elements. These bremsstrahlung continua
are rather flat, the dominant feature being the thermal cutoff at short
wavelengths, where the gratings on board {\it Chandra} and XMM-Newton
have small effective areas. Therefore the continuum's thermal cutoff
constrains the highest temperatures existing in a corona, but to far a
lesser extent the actual run of emission measure vs. temperature.

\subsection{Constraints on and modeling of
the emission measure distribution}

The temperature structure of a stellar atmosphere can be derived
from the principles of radiative and hydrostatic equilibrium, the temperature
structure of a magnetically confined plasma can be computed from the
energy equation, if one assumes -- for example -- a static equilibrium.
The difficulty in the latter case is, that, first, this temperature structure 
of an individual coronal feature is virtually independent on the form of the
assumed heating, which in essence is unknown, and
second, that in the stellar case one is very likely looking at the integrated
emission of a large number of individual features. In other words, this
integrated emission has to be described by some distribution function of the
physical parameters characterizing individual coronal features, and again that
distribution function is unknown. In consequence we conclude that the
mathematical form of the differential emission measure distribution is
{\it a priori} unknown.

Which constraints can nevertheless be imposed on the function describing
the temperature structure, $\xi(T)$ ? In the following we formulate three
conditions and discuss their physical and mathematical implications.
Clearly, from the definition of $\xi(T)$ it follows
\begin{equation}
\xi(T) \ge 0\,. \label{co1a}
\end{equation}
This condition looks trivial, but actually
represents a rather strong constraint. Next, the DEM distributions of
individual magnetic loops have (integrable) singularities at the loop
top; if such DEMs are integrated over distribution functions, continuous,
smooth emission measure distributions are found. We therefore assume
that $\xi(T)$ is a smooth function of temperature. And finally, we
know from the many low resolution X-ray data from proportional counters
that the X-ray temperatures cannot extend to arbitrarily high values.
We therefore assume the existence of some maximum temperature
$T_{max}$, above of which no emission measure is present, i.e.,
\begin{equation}
\xi(T_{max}) = 0\,. \label{co1b}
\end{equation}

In the absence of any plausible physical model we assume that the differential emission
measure $\xi(T)$ can be approximated by a sum of Chebyshev polynomials;
we will discuss a physical ansatz in section 5.3. Of course, any other
system of orthogonal polynomials or other functions could also be used.
Chebyshev polynomials have unique normalization properties, and expansions
of the target function into Chebyshev polynomials are useful as long as the
expansion coefficients diminish rapidly with increasing order. Therefore
this approach has also been taken, for example by Lemen et al. (\cite{lem89})
and Stern et al. (\cite{stern95}).
Also, our
available X-ray data are only sensitive to plasma hotter than some temperature
$T_{min}$, since plasma with $T < T_{min}$
emits outside the band pass. Let us then introduce the dimensionless temperature
variable x defined in the closed interval [0,1] through
\begin{equation}
x = \frac {T - T_{min}} {T_{max} - T_{min}}\,. \label{co2}
\end{equation}
Let $T_n$(x) denote the Chebyshev polynomial of order $n$, satisfying the
following boundary conditions ($n = 0,1,...$):
\begin{equation}
T_{2n}(0) = (-1)^{n} \ \ \ \ \ T_{2n+1}(0) = 0 \ \ \ \ \ \ \ T_{n}(1) = 1\,. \label{co3}
\end{equation}
We seek the differential emission measure distribution
$\xi (x)$ and approximate it through a sum of $M$ Chebyshev polynomials via
\begin{equation}
\xi (x) = \sum_{i=0}^{M} a_i \ T_i(x)\,, \label{co4}
\end{equation}
with coefficients $a_i$, $i=0\ldots M$ to be determined from the
data and the boundary condition
\begin{equation}
\xi(1) = 0\,, \label{co4a}
\end{equation}
which is equivalent with Eq.~\ref{co1b}. Since we are primarily interested in
the shape of the emission measure
distribution (and not in its normalization) we further impose the constraint:
\begin{equation}
\int_{0}^{1} dx \ \xi(x) = 1\,. \label{co4b}
\end{equation}

Because of the properties of Chebyshev polynomials (cf. Eq.~\ref{co3})
the boundary and normalization conditions Eqs. \ref{co4a} and \ref{co4b}
translate into two conditions for the coefficients $a_i$, $i=0\ldots M$:
\begin{equation}
\sum_{i=0}^{M} a_i \ \ = \ \ 0 \label{co5}
\end{equation}
and
\begin{equation}
\sum_{i=0}^{M} a_{i} \int_{0}^{1} dx T_{i}(x) = \ \sum_{i=0}^{M} a_{i} I_{i} \ = \ \ 1\,, \label{co6}
\end{equation}
with known coefficients $I_{i}$.
Therefore the number of independent coefficients $a_i$ is $M-2$ and we can
write the coefficients $a_0$ and $a_1$ as
\begin{equation}
a_1 = 2 \ \sum_{i=2}^{M} (I_{i}- 1)a_{i} - 2 \label{co6a}
\end{equation}
and
\begin{equation}
a_0 = \  \sum_{i=2}^{M} (1- 2 \times I_{i})a_{i} - 2\,. \label{co6b}
\end{equation}
Consider now a specific spectral line denoted by some index $j$ with line
cooling function $\Lambda_j$(x). The measured flux $f_j$ of that line
is given by
\begin{eqnarray}
\label{co7}
f_j & = &\frac {e^{-\sigma _j N_H}} {4 \pi d^2} (T_{max}-T_{min})\ A_X
\int_0^1 dx \Lambda_j(x) \xi (x)\\
&&\nonumber\\
&=&\frac {e^{-\sigma _j N_H}} {4 \pi d^2} (T_{max}-T_{min})\ A_X
\sum_{i=0}^M a_i \int_0^1 dx \Lambda_j(x) T_i(x)\,,\nonumber
\end{eqnarray}

where $d$ denotes the distance to the star, $N_H$ the hydrogen
column density, and $\sigma_j$ the effective absorption cross section per
hydrogen atom at the line's energy. With the abbreviation
\begin{equation}
L_{ji} = \int_0^1 dx \Lambda_j(x) T_i(x) \label{co7a}
\end{equation}
as Chebyshev line contribution coefficients we can write
\begin{equation}
f_j = \frac {e^{-\sigma _j N_H}} {4 \pi d^2} (T_{max}-T_{min})\ A_X
\sum_{i=0}^{M} a_i L_{ji}
\label{co8}
\end{equation}
and one needs to determine $M+1$ coefficients $a_i$, $i=0\ldots M$ from the
data.

\subsection{Differential emission measure reconstruction from lines}

A given line flux $f_j$ depends both on temperature and abundance
(cf. Eq.~\ref{co8}), but the ratio of two emission lines from an ion of the same
atomic species is clearly independent of the specific abundance of the chosen
element. In order to distinguish between abundance and temperature effects on
the differential emission measure distribution one should therefore work
only with line ratios from the same elements (or only with lines from the
same element, which is feasible only for iron). If we let the index $z$ denote
an emission line (or possibly a sum of emission lines of a given element)
in the numerator and analogously the index $n$ the line
in the denominator, we can write for the expected (abundance-independent)
line ratio $\rho_{zn}$
\begin{equation}
\rho_{zn} = \frac {f_z} {f_n} =
\frac {\sum_{i=0}^{M} a_i L_{zi}} {\sum_{i=0}^{M} a_i L_{ni}}\,. \label{ld1}
\end{equation}
Note that the idea of using line ratios is not new; for example,
McIntosh (\cite{mci2000}) has used line ratios in an attempt to determine
the emission measure distribution of features in the solar corona,
and Fludra and Schmelz (\cite{fl95}) have used a similar approach in their
analysis of X-ray flare spectra obtained with the BCS. In our context the great
advantage of using line ratios lies in the fact that in this fashion
the helium-like and hydrogen-like ions of many atomic species can be
used. These emission lines tend to be strongest and the most easily detectable
ones in any stellar corona. For the lighter elements like carbon, nitrogen,
and oxygen these are in fact the only detectable lines in the X-ray range, and
furthermore, atomic physics uncertainties should be smallest for such lines.

We use the double index $(z,n)$ to denote a specific line ratio
and the expressions $Z_{zi}$ and $N_{ni}$ to denote
the line contribution coefficients from the numerator and denominator lines
entering the ratio $(z,n)$, respectively. Given a set $N$ of measured line ratios
$r_{z,n}$ and errors $\sigma_{z,n}$, all of which are derived from emission
lines of the same atomic species, we can determine the differential emission
measure distribution by minimizing the test statistic $\chi^2$ defined as
\begin{equation}
\chi ^2 = \sum_{(z,n)} \frac {\left(\frac {\sum_{i=0}^{M} a_i Z_{zi}}
{\sum_{i=0}^{M} a_i N_{ni}} - r_{(z,n)}\right)^2}{{\sigma_{(z,n)}}^2} \label{ld2}
\end{equation}
with respect to the expansion coefficients $a_i$ for a given set of measured
line ratios and an arbitrary number $M$ of expansion coefficients.
Some care is required since these coefficients are not independent; we use
the boundary conditions Eqs. \ref{co6a} and \ref{co6b} to express the
coefficients $a_0$ and $a_1$ in terms of the remaining coefficients
$a_i, i=2 \dots M$. We further note that multiplying any set of
coefficients $a_i$ by some common scale factor $\Gamma$ leaves
all line ratios $\rho _{(z,n)}$
invariant. As a consequence we can express the test statistic $\chi^2$
with the known parameters $C_{zi}$ for the numerator and $D_{ni}$ for the
denominator in terms of the independent coefficients $a_i,
i=2\ldots M$ as
\begin{equation}
\chi ^2 = \sum_{(z,n)} \frac {\left(\frac {\sum_{i=2}^{M} a_i D_{zi}}
{\sum_{i=2}^{M} a_i C_{ni}} - r_{(z,n)}\right)^2}{{\sigma_{(z,n)}}^2}\,. \label{ld7}
\end{equation}
For the latter expression derivatives can be computed analytically and
minimization is straightforward. In order to ensure that the positivity
constraint Eq.~\ref{co1a} is satisfied, we introduce a penalty function
with large positive values for $\chi^2$ once $\xi(x)$ becomes negative at
any point in the interval $x\in\,$[0,1].

\subsection{Differential emission measure reconstruction from continuum}

The reconstruction of a differential emission measure distribution from
a set of continuum measurements $c_j$ with associated errors $\sigma _j$
follows in analogy to the line treatment, except that obviously the
appropriate continuum contribution functions must be chosen. In general
the temperature sensitivity of the X-ray continuum is much less pronounced
than that of individual emission lines. The most striking feature of the
bremsstrahlung spectra is the exponential cutoff at short wavelengths
resulting from the exponential decay in the number of available
electrons at some given plasma temperature $T$, while at lower temperatures
recombination and two photon continuum can dominate at specific wavelength
bands (see discussion by Mewe et al. \cite{mewe86}).
We also carried out test calculations of
continua with ``reasonable'' differential emission measure distributions
and found the spectral shape of these continua quite insensitive to
variations in the parameters of the differential emission measure
distribution. In our fits we therefore used only a small number (3) of
continuum bands. 

\section{Which lines to use ?}

\begin{figure}
\resizebox{\hsize}{!}{\includegraphics{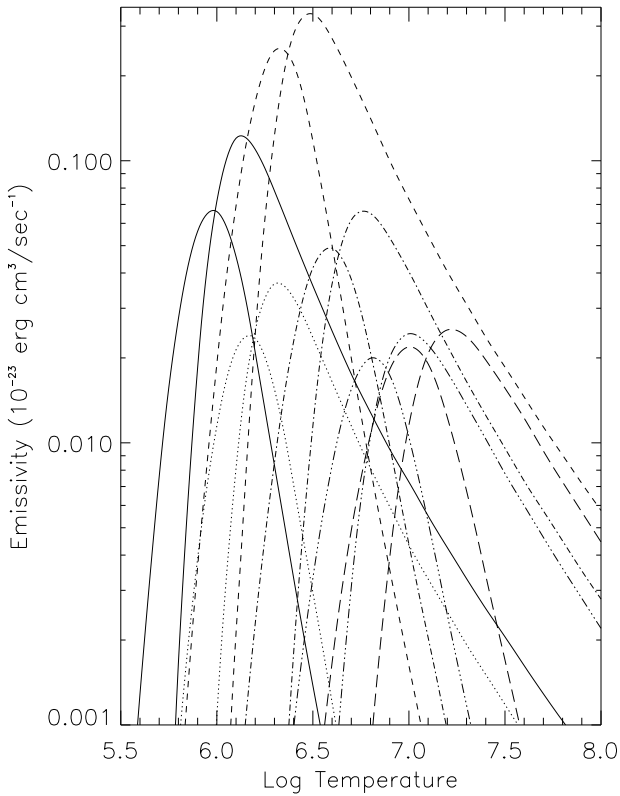}}
\caption[]{\label{em_theo}Theoretical line fluxes for a cosmic abundance
plasma for H-like and He-like resonance lines for C (solid lines),
N (dotted lines), O (short-dashed lines), Ne (dash-dotted lines),
Mg (dash-triple dotted lines), and Si (long-dashed lines) as calculated
with the MEKAL code. Note, that for all elements the Ly$_{\alpha}$ lines are
stronger than the He-like resonance lines.}
\end{figure}

An inspection of Eq.~\ref{l6} shows that the temperature dependence of the
power emitted in a given spectral line is determined by two factors, i.e.,
first, the ionization equilibrium of the atomic species considered and
second, by the electron excitation rate. The ionization equilibrium will
in general be such that the fractional ionization of a given stage of
ionization, say, O\,{\sc vii}, peaks at some temperature, and at lower (higher)
temperatures the predominant ionization will be lower (higher) and the
emitted line flux will correspondingly change. As to the electron excitation
rate, in general for excitation of a given line a certain threshold energy
is required, and all electrons above this threshold value will be able
to perform atomic excitations. Therefore, the excitation rates will in general
increase with increasing temperature but eventually level out. Therefore, the
emissivity of a given line of some species, say O\,{\sc vii}, will peak
at some temperature $T_{max}$.
Note that the line of the hydrogen-like ions are
broader (with high temperature tails) than the corresponding lines from
the helium-like ions, an effect caused by the ionization equilibrium.
In order to illustrate this behavior we plot in
Fig.~\ref{em_theo} the line emissivity (per unit emission measure) as a
function of temperature for the Ly$_{\alpha}$ and He-like resonance lines
(i.e., the transition $^1P_1$ - $^1S_0$) for the elements carbon, nitrogen,
oxygen, neon, magnesium, and silicon according to the calculation by
Mewe et al. (1985). As obvious from Fig.~\ref{em_theo}, these
lines very nicely sample the temperature range between $\approx$ 10$^6$ and
$\approx$ 2$\times$ 10$^7$\,K. These lines belong to
the strongest lines of these elements in this temperature range and
are easily observable in {\it Chandra} LETGS spectra.

\begin{figure}
\resizebox{\hsize}{!}{\includegraphics{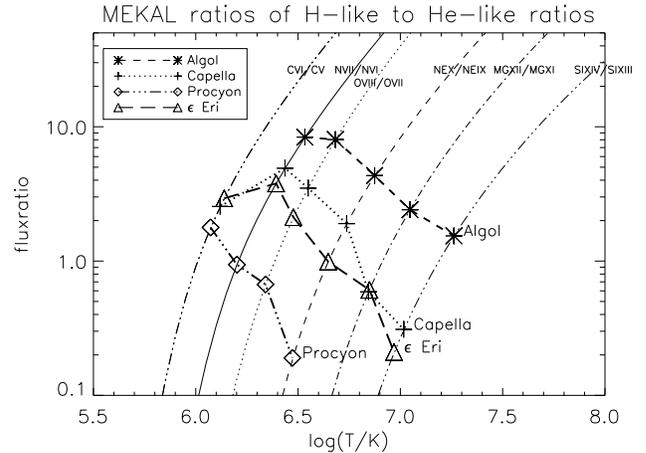}}
\caption[]{\label{lyrats}Theoretical line flux ratios of H-like by He-like
resonance lines for C, N, O, Ne, Mg, and Si in comparison with the measured
ratios for Algol, Capella, and Procyon.}
\end{figure}

If one now considers line ratios from the same chemical element but
from two adjacent stages of
ionization, say from O\,{\sc vii} and O\,{\sc viii}, the lower ionization
stage line will dominate at lower temperature and vice versa, and the line
ratio will monotonically increase with temperature. In Fig.~\ref{lyrats} we
show the ratio $\rho$ of the energy flux in the Ly$_{\alpha}$ line divided
by the He-like r line as a function of temperature $T$. As is clear from
Fig.~\ref{lyrats}, the line ratios $\rho$ do indeed increase monotonically with
temperature. The temperatures at which $\rho$ is unity for a given
atomic species increase with increasing atomic mass (and nuclear charge)
reflecting the fact that more and more energy is required to establish,
say, the He-like stage of ionization. $\rho$ increases by $\approx$ 100
for a temperature increase of $\approx$ 0.5 dex, thus $\rho$ is rather
temperature sensitive. Therefore a given measured line ratio can be (uniquely)
converted into a temperature, however, different line ratios will in
general result in different temperatures. These temperatures must not
be interpreted as "isothermal" temperatures, but rather as "moment" temperatures
since they depend on the differential emission measure distribution
(a stellar property) and the lines' emissivity functions (an atomic property).

%
%
%
%
%

How do these curves compare to observations? We included the measured data
points for the stars Algol (high activity), $\epsilon$ Eri and Capella
(intermediate activity), and Procyon (low activity) taken from Ness et al.
\cite{ness10} who gathered together line ratios of hydrogen-like/helium-like
line intensities for all ions measurable with the LETGS.
As can be seen from Fig.~\ref{lyrats}
the largest value for $\rho_{Si}$ is obtained for Algol ($\rho_{Si}$ = 1.54),
while for the low-activity stars no lines from H-like or He-like silicon
and magnesium are observed. For Algol the $\rho$-values increase with
decreasing atomic number reaching $\rho_{N}$ = 8.37 for nitrogen; no lines
from carbon are observed. For Capella $\rho_{Si}$ and $\rho_{Mg}$ are below
unity, then $\rho$ increases to $\rho_{N}$ = 4.91 for nitrogen while
$\rho_{C}$ is lower. For the low-activity star Procyon all $\rho$-values
except for carbon ($\rho_{C}$ = 1.78) are below unity; the neon lines
are observed only quite weakly. So clearly the available data suggest
large differences in the emission measure distribution for the sample stars.

We will base our differential emission measure reconstruction solely on
those lines. The disadvantage of using those lines, i.e., their formation over
relatively broad temperature range, is in our opinion more than compensated
by a number of advantages: First of all, the atomic physics of hydrogen- and
helium-like ions is much simpler than that of more complicated ions.
Second, these lines are among the strongest lines; they can therefore
be measured in a large sample of stars (cf. Ness et al. \cite{ness10}) and the
DEM reconstructions of different stars can be compared with each other
since they are computed in the same fashion. And third, as we will show
below, these lines are very likely to already contain most of the temperature
information contained in stellar coronae. It is important to realize that
more line ratios do not necessarily provide more information on the
temperature structure; additional line ratios may either contain no or
no new information or may provide conflicting information. For example,
McIntosh (\cite{mci2000}) gives the measured and calculated line ratios
used in his differential emission reconstruction (cf. his Fig. 3b), which
deviate by almost an order of magnitude in the worst case.

\section{Which continuum to use ?}

In order to fix the overall normalization we use both
the measured shape of the continuum and the absolute
level of the observed continuum
radiation. The first problem to be solved - a problem very familiar
to optical astronomers - is the correct placement of the continuum.
While strong lines can be recognized easily, the sum of weak lines,
each of which remains undetected, can in principle produce a
``pseudo-continuum''. Since specifically the LETGS covers such a
large band pass it appears unlikely that over the whole instrument band pass
from 5\,\AA\ - 170\,\AA\ such a ``pseudo-continuum'' is produced, while
in narrower spectral bands this may well be the case. In order to isolate
the continuum we use a median filter in the following way: In a predefined
wavelength region - typically we use 0.5\,\AA\ - we calculate the median
and use this value as the characteristic continuum level at that
particular wavelength. Clearly, if too many lines are located in the
wavelength bin considered, the thus derived continuum level is too high.
This is specifically the case in the rather crowded region between 9\,\AA\
and 18\,\AA, where it is next to impossible to reliably place any continuum.
Fortunately, other spectral regions are far less crowded and do allow a
rather reliable continuum placement.

\section {A worked out example: Algol}

An 80\,ksec observation of the eclipsing binary Algol has been
carried out with the LETGS on board {\it Chandra}; the recorded data
set and an analysis of the He-like and H-like lines has been presented
by Ness et al. (\cite{ness_alg}), while Schmitt and Ness (\cite{schm2002})
discuss the carbon and nitrogen abundances of Algol and other giants.
Here we focus on the determination of
the differential emission measure distribution and elemental abundances;
for a comparison of the coronal spectra of HR1099 and Algol B we refer to
Drake  (\cite{drake03}).

\subsection {DEM modeling with Chebyshev Polynomials}

\subsubsection{Temperature structure}

Algol's X-ray emission is very strong and except for carbon all
H-like and He-like lines from N, O, Ne, Mg, and Si were detected;
the failure to detect carbon lines in the X-ray spectrum of Algol
is model-independent and due to a nitrogen enrichment of CNO-cycle
processed material (Schmitt \& Ness \cite{schm2002}). Thus unfortunately
no information is available in the lower temperature range of the
emission measure distribution from hydrogen- or helium-like lines.
The flux ratio between the Si Ly$_{\alpha}$ and He-like resonance line
exceeds unity indicating that the peak of the emission measure distribution
is beyond 10\,MK. We therefore consider four values of T$_{max}$, i.e.,
20\,MK, 30\,MK, 40\,MK, and 50\,MK. We first considered the simplest
possible combination of Chebyshev polynomials with M=4, constrained to
yielding a positive emission measure distribution. In order to prevent
negative emission measures we introduced a penalty function resulting in
large values whenever the reconstructed emission measure distribution has
negative values in the
interval between T$_{min}$ and T$_{max}$; we chose T$_{min} = 4 \times
10^5$\,K, and note that our results are not very sensitive to the specific
choice of T$_{min}$. We specifically point out that the line with the coolest
peak formation temperature is the resonance line of N\,{\sc vi} with a peak
formation temperature of $\approx 10^{6.1}$\,K (cf. Fig.~\ref{em_theo}) and
with our {\it Chandra} LETGS data we have little information on the emission
measure distribution below 10$^6$. For each permitted emission measure
distribution the resulting line ratios of N, O, Ne, Mg, and Si were computed
and compared to the observed line ratios via the $\chi ^2$ test statistics.

In this paper we use the {\it Chianti} software package
(cf. Young et al. \cite{youn03}) to compute
both line and continuum intensities for all modelling of
{\it Chandra} LETGS data. The line intensities were computed
in photon mode, for the continuum the contributions from bremsstrahlung,
recombination continuum, and two photon continuum were separately computed
and added. The ionization equilibrium by Mazzotta et al. (\cite{mazz98})
was used. All calculations were carried out using cosmic abundance
as quoted by Allen (\cite{all73}).  For the relevant elements we specifically used
the values [C/H]= 8.52, [N/H]= 7.96, [O/H]= 8.82, [Ne/H]= 7.92, [Mg/H]= 7.42, [Si/H]= 7.52, and
[Fe/H] = 7.60.  Comparing these abundance values to the most recent values
given by Grevesse and collaborators (\cite{grev98}), we find essentially
identical values for C, N, O and Si, while for the elements Ne, Mg and Fe
Grevesse \& Sauval (\cite{grev98}) give [Ne/H]= 8.08, [Mg/H]= 7.58 and
[Fe/H] = 7.50, i.e., values differing by factors of 1.44 (Ne and Mg) and
0.79 (Fe) respectively.   We only give the multiplicative factors,
by which our derived abundance values differ from the ones quoted by
Allen (\cite{all73}); in order to convert to the ones quoted by  
Grevesse \& Sauval (\cite{grev98}), the He and Mg abundances must by multiplied
by 0.69, the Fe abundance by 1.26.

\begin{figure}
\resizebox{\hsize}{!}{\includegraphics{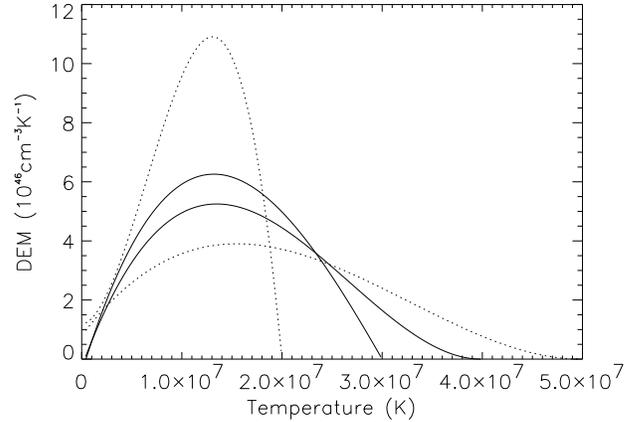}}
\caption[]{\label{figdem1} Best fit differential emission measure distributions
with fourth-order Chebyshev polynomials and temperatures $T_{max}$ = 20\,MK,
30\,MK, 40\,MK, and 50\,MK; only the choices $T_{max}$ = 30\,MK and
$T_{max}$ = 40\,MK yield acceptable fits to the {\it Chandra} LETGS data.
}
\end{figure}

\begin{table*}
\caption[ ]{\label{res1tab} {\sc Line ratio fit results for DEM models based on
Chebyshev polynomials.}
The numbers given in parentheses are ($T_{max}$, $M$).  For all analysed
atomic species we give the observed line ratios with their errors as
well as the modelled line ratios for the various models considered.
The last row gives the values of the $\chi ^2$ test statistics
for the respective models.}
\renewcommand{\arraystretch}{1.0}
\begin{tabular}{|c|c|c|c|c|c|c|}
\hline
Element& Observed  &Modelled  &Modelled  &Modelled  &Modelled  &Modelled  \cr
       &ratio/error&line ratio&line ratio&line ratio&line ratio &line ratio \cr
       &           & (20,4)   & (30,4)   &(40,4)    &(50,7)    &(50,4)  \cr
\hline
N  & 8.38$\pm$1.28 & 8.85 &     9.44     & 9.21 &  9.00 & 6.66\cr
O  & 8.00$\pm$0.71 & 7.37 &     7.02     & 7.04 &  6.84 & 5.94\cr
Ne & 4.46$\pm$0.28 & 4.80 &     4.61     & 4.76 &  4.41 & 4.77\cr
Mg & 2.41$\pm$0.31 & 2.27 &     2.52     & 2.69 &  2.48 & 2.97\cr
Si & 1.55$\pm$0.14 & 1.10 &     1.49     & 1.66 &  1.63 & 1.97\cr
\hline
$\chi^2$&          & 12.45&     3.16     & 4.82 & 3.36  & 12.89\cr
\hline
\end{tabular}
\end{table*}
Our modeling attempts showed that already with the choice of $M=4$ good
fits to the line ratios of the Ly$_{\alpha}$ and He-like resonance lines
can be obtained. In Fig.~\ref{figdem1} we plot the
best fit reconstructed emission measures (for the case $M=4$) for the
peak temperatures $T_{max}=20$\,MK, 30\,MK, 40\,MK, and 50\,MK; both the
curves for $T_{max}= 30$\,MK and 40\,MK yield acceptable fits (solid lines),
while the choices of $T_{max}=20$\,MK and $T_{max}=50$\,MK lead to
unacceptable fits (cf. Tab.~\ref{res1tab}).
Our formal fit results are presented in
Tab.~\ref{res1tab}, where we give for those four best fits the resulting
values of $\chi ^2$ for the line ratio fit as well as the modelled line
ratios.
A maximum temperature of
$T_{max}$ = 20\,MK is simply too low to explain the observed ratio between
Ly$_{\alpha}$ and He-like resonance line for silicon. On the other hand, for
the model with $T_{max}$ = 50\,MK and M=4 too much emission measure is located
at high temperatures leading to an increase in $\chi ^2$.
In all fits we also checked for the goodness of fit to the continuum;
in general,
fits with higher temperature result in better continuum fits than
lower temperatures.
\begin{figure}
\resizebox{\hsize}{!}{\includegraphics{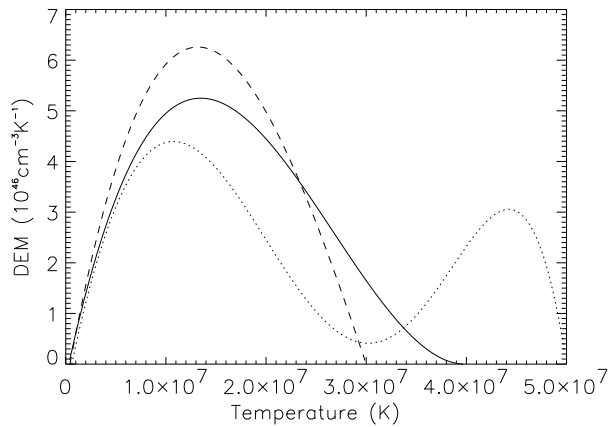}}
\caption[]{\label{figdem2} Best fit differential emission measure distributions
with Chebyshev polynomials with ($T_{max}=30$\,MK, M=4; dashed curve),
($T_{max}=40$\,MK, M=4; solid curve), and ($T_{max}=50$\,MK, M=7;
dotted curve).
All parameter choices yield acceptable fits to the {\it Chandra} LETGS line
ratio data, the higher temperature yield better continuum values.
}
\end{figure}

Clearly, the restriction to the first four Chebyshev polynomials
results in rather
simple emission measure distributions, which, however, provide good line ratio
fits for the correct choices of $T_{max}$. How unique are the derived
emission measure distributions ? In order to assess this issue we
introduced higher order Chebyshev polynomials, which result in more
complicated emission measure distributions and in improved fits.
In Fig.~\ref{figdem2} we plot the best fit reconstructed
emission measures for the cases ($T_{max} = 30$\,MK,M = 4),
($T_{max} = 40$\,MK,M = 4) and ($T_{max} = 50$\,MK,M = 7)
as a function of temperature; the resulting $\chi^2$ values are given
in Tab.~\ref{res1tab}. As can be seen from Tab.~\ref{res1tab}, all fits
have similar goodness of fit parameters, but the resulting DEM curves are
quite different. In particular, assuming $T_{max} = 50$\,MK, leads
to a bimodal emission measure distribution with a second peak at 45\,MK,
corresponding to a cutoff energy of about 3\,keV.
The presence of such a peak in the
emission measure distribution can be readily recognized from the high energy
continuum emitted by hot plasma. However, the LETGS is not particularly
sensitive in that wavelength range. For XMM-Newton data, for example,
simultaneously taken CCD spectra at higher energies would yield
important constraints at the high temperature end of the DEM distribution
which are not provided by the LETGS.
From a statistical point of view the improvements in fit quality are
so small that
they do not warrant the introduction of additional degrees of freedom.

\begin{figure}
\resizebox{\hsize}{!}{\includegraphics{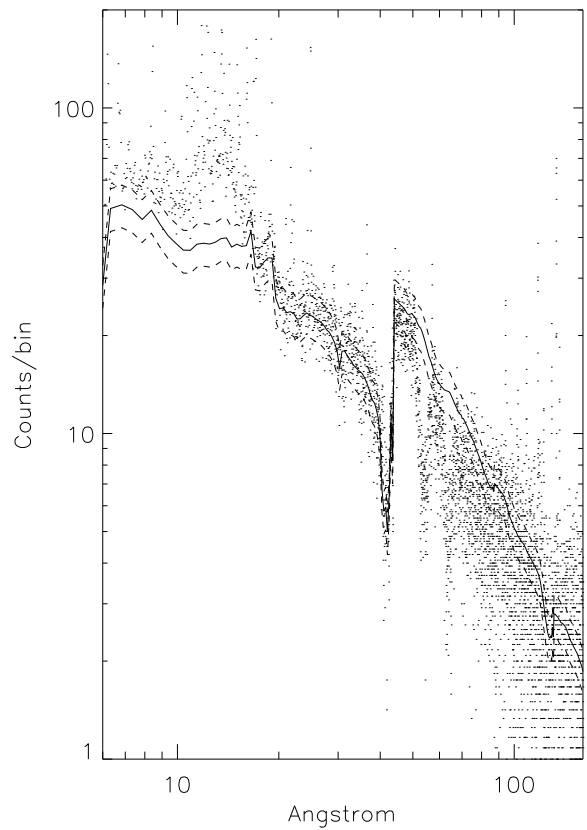}}
\caption[]{\label{cfit} {\it Chandra} LETGS spectrum for Algol binned
in wavelength bins with a width of TBD \AA (dots) and best fit
differential emission measure model continuum for $M=4$ and $T_{max} = 40$\,MK
(solid line); dashed curves indicate a 15\% systematic uncertainty in
effective instrument areas. Note that the continuum has been calculated
from the abundances actually derived for Algol.}
\end{figure}

In order to demonstrate the overall goodness of fit of our continuum models we
plot in Fig.~\ref{cfit} the observed {\it Chandra} LETGS for Algol (with
a resolution of 0.03\,\AA; small
dots) and the best fit continuum model (solid line) for the case
$T_{max}=40$\,MK and $M=4$. 
We emphasize that in fixing the normalization we attempted only modeling the
continuum, but not the individual lines; the continuum modeling includes
higher orders up to order ten.  The continuum was computed with the
specific set of abundances calculated for Algol, but no iterative 
(re-)determination of abundances was performed, since the changes in the
continuum level are of the order of a few percent at most. 
The continuum was fitted in the range
19\,\AA\ - 40\,\AA\ and 70\,\AA\ - 105\,\AA.
Fig.~\ref{cfit} shows that the continuum
is well fitted at short wavelengths and in the wavelength range 80 -
100\,\AA. The fit is extremely poor in the wavelength range between
8 - 18\,\AA; this is hardly surprising since numerous emission lines are
located (cf. Fig.~1 in Ness et al. \cite{ness_alg}) in that wavelength region.
Our fit describes quite well the carbon edge at 44\,\AA. Our fits behave badly
near $\approx$ 53\,\AA\ and 63\,\AA, where the LETGS spectra show two dips,
which are instrumental and caused by gaps in the HRC-S microchannel plates.
At long wavelengths our continuum fit describes the observed data, but is
somewhat high. We checked our analysis procedures on the public 
{\it Chandra} LETGS data (500 ksec) of the isolated neutron star RXJ1856.5-3754
and found very good agreement between our results and those published in the
literature. 
We have at present no satisfactory explanation for the discrepancy at longer wavelengths. 
On the one hand our choice of median filtering has some bias
towards higher values because the filtering procedure essentially regards
spectral lines as background fluctuations. On the other hand errors in the
instrument calibration can also not be ruled out; calibrations in the EUV
are notoriously difficult and an error in the relative calibration between the
long- and short wavelength region will also help to reduce the observed discrepancy.
An even larger absorption column towards Algol would also improve the fit.

\subsubsection{Abundances}
With the reconstructed abundance independent emission measure
distributions (cf. Fig.~\ref{figdem1}) and the measured line fluxes
we can now calculate elemental abundances. Specifically, we can
calculate for each observed spectral line that value of abundance
which forces agreement between predicted and observed line fluxes.
Obviously, the abundances for a given chemical element will depend on the
specific lines used and any errors in the atomic parameters and
temperature structure will propagate into the derived abundances;
our results for various lines are summarized in Tab.~\ref{abun}.
All abundance values quoted are relative to the values [C/H]= 8.52, 
[N/H]= 7.96, [O/H]= 8.82, [Ne/H]= 7.92, [Mg/H]= 7.42, [Si/H]= 7.52, and
[Fe/H] = 7.60.
An inspection of Tab.~\ref{abun} shows that H- and He-like lines
give consistent abundance values; this is of course no surprise, since
the adopted differential emission measure distribution fits
the line ratios very well. The upper limit derived from the absence
of the C\,{\sc vi} Ly$_{\alpha}$ line implies that the carbon abundance
(relative to solar) is at the level of less than 10 percent, i.e.,
most of the carbon has been depleted. No iron lines have been
used for the DEM reconstruction. Inspection of Tab.~\ref{abun}
shows that the nitrogen abundances for accpetable models lie between
1.9-2.9, the oxygen abundances between 0.25-0.36,
the neon abundances between 0.95-1.34, the magnesium abundances
between 0.48-0.68, and those of silicon between 0.42-0.61.
The Fe abundances (for the best fit DEM models) vary from 0.10 - 0.49
depending on the lines chosen. These discrepancies are of course
worrisome, but they are small compared to the discrepancies found
in solar work (cf. McIntosh \cite{mci2000}). Also, some Fe lines
consistently yield abundance values inconsistent with those derived
from other lines. For example, the Fe\,{\sc xvii} 17.07\,\AA\ line yields
larger values than the neighboring 15\,\AA\ line, and the
Fe\,{\sc xix} 101.55\,\AA\ line gives much higher values than the other
Fe XUV lines, indicating systematic problems with those lines.
Tab.~\ref{abun}
also reveals that the Fe abundances derived from XUV lines (with the
exception of Fe\,{\sc xix} 101.55\,\AA )are 
in general lower than those of X-ray lines. Whether this is due to
atomic physics or
instrumental problems is difficult to say at present. The errors
in the effective area calibration of the LETGS are thought to be of
the order $\pm$\,15\,\% at most; the magnitude of this effect appears
to be too small to account for the observed discrepancies.  Another
issue is the value of the interstellar absorption column density;
for our analysis we used the value of 5$\times 10^{-18}$\,cm$^{-3}$,
and higher values would
reduce, smaller values increase the discrepancy between model and
observations.

\begin{table*}
\caption[ ]{\label{abun}Abundance determinations for DEM models based on
Chebyshev polynomials; listed are the elements and lines used (first two
columns), the measured number of counts, the modelled number of counts as well
as the abundance
relative to the Allen (\cite{all73}) abundance values for the various models 
characterized by
number of polynomials and peak temperature. Last row indicates whether fits
are acceptable ``{\it acc}'' or unacceptable ``{\it not acc}''.}

\renewcommand{\arraystretch}{1.2}
\begin{tabular}{|l|rr|rr|rr|rr|rr|rr|}
\hline
Ion       & Line       & Obs.     & Mod.  &(7,50)& Mod.  &(4,30)& Mod.  &(4,40) & Mod. & (4,50) & Mod. & (4,20)\cr

          &    (\AA)   &          &       &          &          &         &           &        &          &        &          &       \cr
\hline
C\,{\sc vi} & 33.74    &  $<$ 60  & 588.2 & $<$ 0.10 & 763.1 &  $<$ 0.08 & 660.5 & $<$ 0.09 & 659.8 & $<$ 0.09 & 908.1 & $<$ 0.05 \cr
N\,{\sc vi} & 24.785   & 1119.1   & 416.9 & 2.68 & 567.4 & 1.97 & 501.1 & 2.23 & 467.7 & 2.39 & 673.0  & 1.71 \cr
N\,{\sc vii} &  28.79  & 141.3    & 48.6  & 2.91 & 63.3  & 2.23 & 57.0  & 2.48 & 54.4  & 2.59 & 76.9   & 1.84 \cr
O\,{\sc viii} & 18.97  & 2883.0   & 7925.2& 0.36 &10289.0& 0.28 & 9057.4& 0.32 & 8890.1& 0.32 & 11749.0& 0.25 \cr
O\,{\sc vii} & 21.6    & 262.5    & 836.7 & 0.31 &1059.7 & 0.25 & 930.1 & 0.28 & 938.5 & 0.28 & 1251.5 & 0.21 \cr
Ne\,{\sc x} & 12.138   & 2481.5   & 1956.4& 1.27 &2602.2 & 0.95 & 2300.1& 1.09 & 2194.7& 1.13 & 2996.8 & 0.83 \cr
Ne\,{\sc ix} & 13.45   & 665.0    & 497.8 & 1.34 &633.5  & 1.05 & 542.4 & 1.23 & 558.4 & 1.19 & 710.7  & 0.94 \cr
Mg\,{\sc xii} & 8.43   & 578.0    & 878.0 & 0.66 &1196.3 & 0.48 & 1086.1& 0.53 & 985.0 & 0.59 & 1406.1 & 0.41 \cr
Mg\,{\sc xi} & 9.17    & 224.1    & 329.3 & 0.68 &439.7  & 0.51 & 374.9 & 0.60 & 369.4 & 0.61 & 514.6  & 0.44 \cr
Si\,{\sc xiv} & 6.15   & 658.3    & 1135.8& 0.58 &1504.8 & 0.44 &1452.2 & 0.45 & 1274.1& 0.52 & 1784.5 & 0.37 \cr
Si\,{\sc xiii} & 6.65  & 480.7    & 784.7 & 0.61 &1132.1 & 0.42 & 983.0 & 0.49 & 880.3 & 0.55 & 1618.6 & 0.41 \cr
Fe\,{\sc xvii} & 15.01 & 1018.4   & 5322.6& 0.19 &6740.0 & 0.15 & 5714.8& 0.18 & 5970.6& 0.17 & 7286.7 & 0.14 \cr
Fe\,{\sc xvii} & 15.27 & 364.7    & 1286.1& 0.28 &1628.1 & 0.22 & 1380.6& 0.26 & 1442.6& 0.25 & 1760.1 & 0.21 \cr
Fe\,{\sc xvii} & 17.07 & 1124.5   & 2278.8& 0.49 &2879.7 & 0.39 & 2443.5& 0.46 & 2556.2& 0.44 & 3109.9 & 0.36 \cr
Fe\,{\sc xvii} & 93.92 & 258.0    & 2023.9& 0.13 &2599.7 & 0.10 & 2196.2& 0.12 & 2270.3& 0.11 & 2857.9 & 0.09 \cr
Fe\,{\sc xix}  &108.37 & 203.5    & 1193.6& 0.17 &1571.0 & 0.13 & 1323.1& 0.15 & 1339.0& 0.15 & 1817.1 & 0.11 \cr
Fe\,{\sc xix}  &101.55 & 180.9    & 444.9 & 0.41 &585.5  & 0.31& 493.2 & 0.37  & 499.0 & 0.36 & 677.3  & 0.25 \cr
Fe\,{\sc xxi} & 128.73 & 266.32   & 1914.5& 0.14 &2725.2& 0.10& 2300.5& 0.12 &2147.8 & 0.12 & 3730.5 & 0.07 \cr
Fe\,{\sc xxiii}&132.85 & 1178.6   & 5561.9& 0.21 &8505.0& 0.14& 7508.5& 0.16 &6250.8 & 0.19 &11021.3 & 0.11 \cr
\hline
               &       &          &         & acc &       & acc &         & acc &    & not acc &  & not acc\cr
\hline
\end{tabular}

\begin{flushleft}
\renewcommand{\arraystretch}{1}
\end{flushleft}
\end{table*}

\subsection{Temperature component analysis}

\subsubsection{Temperature structure}

It is of course also possible to carry out the analysis of the line ratio data
in a more traditional way using individual, discrete temperature
components. In this approach one uses $L$ independent spectral components with
temperatures $T_l$, $l = 1 \dots L$ and emission measures $A_l$. One clearly
ought to demand
\begin{equation}
A_l \ge 0\,, \label{poscon_disc}
\end{equation}
and, since we are using line ratios, any solution
is only determined up to a multiplicative factor which needs to be fixed by
the continuum. We use the constraint
\begin{equation}
$$ \sum_{l=1}^L A_l = 1 $$
\end{equation}
and therefore a solution with $L$ spectral components has $2L-1$ fit parameters.
In Tab.~\ref{abunconv} we present the results of our analysis of the available
line-ratios for Algol using individual temperature components; we list the
best fit temperatures and amplitudes as well as the goodness of fit parameter
$\chi ^2$. An inspection of Tab.~\ref{abunconv} shows that a good fit to the
observed line ratios measured for the Ly$_{\alpha}$- and He-like resonance
lines for N, O, Ne, Mg, and Si requires four spectral components with a total
of seven adjustable parameters. Descriptions with fewer components and fewer
adjustable parameters cannot explain the observed line ratios; the test
statistic $\chi^2$ depends - not surprisingly - sensitively on the number of
spectral components, but three components, yielding a test statistic
$\chi_{min}^2=17.15$, are just not sufficient, while the introduction of a
fourth and fifth component reduces the level of $\chi^2$ to below unity. The
necessity to introduce 7 parameters in order to fit 5 data points is a
nightmare for every statistician and one wonders about the effective number of
degrees of freedom. Presumably these problems have to do with the positivity
constraint Eq.~\ref{poscon_disc}, which severely limits the available solution space.
Unfortunately, the resulting best fit temperature quadruple is not unique.
If one accepts a limit of $\chi_{lim}^2 \le 6$ as a threshold below which
solutions are accepted uncertainties of up to 0.2 dex in Log $T$ result.

\subsubsection{Abundances}

One expects that the derived abundances depend sensitively on the choice
of the adopted temperature components if one is working with a small number
of temperature components. These expectations are verified by an inspection
of Tab.~\ref{abunconv}, where we list the abundances derived from our
temperature fits. In Tab.~\ref{abunconv} we list the derived elemental
abundances for each discrete temperature solution for three, four, and five
individual components (list in the first row). The depletion of carbon is
recognized independent of the chosen number of temperature components, while
abundances for individual elements and lines can but must not vary considerably.
For example, the O abundance derived from O\,{\sc viii} 18.97\,\AA\ changes
from 0.19 - 0.20, i.e., is essentially model independent, while the iron
abundance derived from Fe\,{\sc xx} 101.55\,\AA\ varies from 0.06 to 0.24 in
an extremely model dependent way.

If we now compare the abundances from the simplest acceptable
discrete temperature
component model (i.e., 4-T) to those derived from the simplest continuous
emission measure distribution model (4, 40) we find that the nitrogen and
oxygen abundances differ by a factor of $\approx$ 2, the discrete temperature
component abundances being lower. For the neon, magnesium, and silicon
abundances the situation appears similar with the continuous temperature distribution
abundances being higher, while for iron the two sets of abundances agree
reasonably well the main difference coming from the lines used for abundance
analysis.

\begin{table}
\caption[ ]{\label{abunconv}Abundance determinations for discrete
temperature component models; listed are the elements and lines used (first two
columns), the model parameters in parentheses indicating number of temperature
components and value of $\chi^2$ as well as the individually derived abundances
relative to cosmic abundances. Last row indicates whether model gives acceptable
``{\it acc}'' or unacceptable fits ``{\it not acc}''.}
\renewcommand{\arraystretch}{1.2}
{\scriptsize
\begin{tabular}{|l|rrrr|}
\hline
Ion & Line (\AA) & (3,9.1) &  (4,2.26) &  (5,0.83)\cr
\hline
C\,{\sc vi}  & 33.74  & $<$ 0.06  & $<$ 0.06 &  $<$ 0.06\cr
N\,{\sc vi}  & 28.79   & 1.54 & 1.41 & 1.45 \cr
N\,{\sc vii} & 24.785  & 1.35 & 1.41 & 1.45 \cr
O\,{\sc vii} & 21.6    & 0.19 & 0.17 & 0.19 \cr
O\,{\sc viii} & 18.97  & 0.20 & 0.19 & 0.19 \cr
Ne\,{\sc ix} & 13.45   & 0.68 & 0.77 & 0.74 \cr
Ne\,{\sc x} & 12.138   & 0.58 & 0.69 & 0.69\cr
Mg\,{\sc xi} & 9.17    & 0.28 & 0.33 & 0.33 \cr
Mg\,{\sc xii} & 8.43   & 0.39 & 0.36 & 0.37 \cr
Si\,{\sc xiii} & 6.65  & 0.33 & 0.35 & 0.36\cr
Si\,{\sc xiv} & 6.15   & 0.34 & 0.34 & 0.34 \cr
Fe\,{\sc xvii} & 15.01 & 0.06 & 0.09 & 0.09 \cr
Fe\,{\sc xvii} & 15.27 & 0.09 & 0.13 & 0.13 \cr
Fe\,{\sc xvii} & 17.07 & 0.16 & 0.24 & 0.23\cr
Fe\,{\sc xvii} & 93.92 & 0.05 & 0.05 & 0.05 \cr
Fe\,{\sc xx}  & 101.55 & 0.24 & 0.15 & 0.06\cr
Fe\,{\sc xix}  &108.37 & 0.10 & 0.06 & 0.06\cr
Fe\,{\sc xxi} & 128.73 & 0.16 & 0.14 & 0.13\cr
Fe\,{\sc xxiii}&132.85 & 0.11 & 0.15 & 0.15\cr
\hline
        &       & not acc & acc & acc\cr
\hline
\end{tabular}
}
\begin{flushleft}
\renewcommand{\arraystretch}{1}
\end{flushleft}
\end{table}

\subsection{Towards a physical model}

It should have become clear that neither the modeling approach based on
Chebyshev polynomials nor that based on individual temperature components
includes a great deal of physics other than the theory of hot thermal plasma
emission. Let us consider a ``standard'' solar loop with apex temperature
$T_{max}$ = 2 $\times$ 10$^6$\,K, pressure $p$ = 1\,dyn\,cm$^{-2}$, and loop
half length L = 2 $\times$ 10$^9$\,cm. Such a structure contains a total
emission measure of a few times 10$^{46}$\,cm$^{-3}$. Comparing this estimate
to the total solar coronal emission measure or to the emission measure
observed from stars it is clear that several hundreds and possibly more than
thousand such "standard" loops must be contributing to the emission observed
at any given time. This is amply demonstrated by the thousands of {\it YOHKOH}
and {\it SOHO} images, which show a vast variety of different emission
structures, and only during a stronger flare an individual structure can
dominate the total X-ray output. In a stellar context the situation is less
clear. Again, during a large flare the overall emission is certainly dominated
by the emission from the flare region alone, for the quiescent emission we
assume that a larger number of individual features is responsible for the
observed emission.

\begin{figure}
\resizebox{\hsize}{!}{\includegraphics{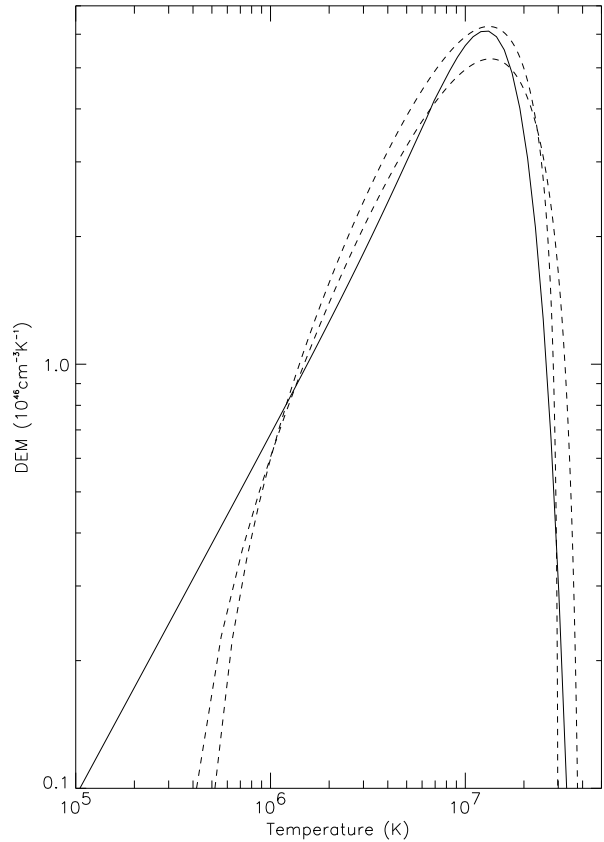}}
\caption[]{\label{comp} Best fit differential emission measure
distribution derived from Eq.~\ref{pm7} (solid curve)
compared to DEM distributions
derived from fourth order Chebyshev polynomials with $T_{max}=$30\,MK
and $T_{max}=$40\,MK.}
\end{figure}

If one assumes that the X-ray emission comes from a sample of individual
constant pressure ``atmospheres'' extending from some lower temperature
$T_0$ to some maximum temperature $T_{max}$, the differential emission
measure distribution $\xi(T)$ of such a loop can be calculated as
(cf. Bray \cite{bray91})
\begin{equation}
\xi(T) = C_{norm}\ p\ T^{\alpha -\gamma/2 - 1/4}
\frac {1} {\sqrt{1-(\frac {T} {T_{max}})^{2-\gamma + \beta}}}. \label{pm1}
\end{equation}
The parameters $\alpha$, $\beta$, and $\gamma$ in Eq.~\ref{pm1} are
the power law coefficients in the laws for radiative cooling, loop heating,
and loop cross section expansion which are assumed to have the form
\begin{equation}
\Lambda (T) = \Lambda_0 T^{\gamma} \ \ \ Q(T) = Q_0 T^{\beta} \ \ \ A(T) = A_0 T^{\alpha}\,,
\label{pm2}
\end{equation}
respectively. Clearly, these assumptions are somewhat idealistic, and in
particular the assumption about the heating law is totally arbitrary, but the
resulting structure is found to be remarkably insensitive to the detailed
form of these coefficients.
The constant $C_{norm}$ in Eq.~\ref{pm1} is given by the expression
\begin{equation}
C_{norm} = A_0 \sqrt{\frac {\kappa_0 (2\alpha + \gamma + 3/2)}
{8 \Lambda_0 k^2}}\,, \label{pm3}
\end{equation}
where $\kappa_0$ denotes the constant in the heat conduction law.
The singularity at the loop top, where $T=T_{max}$, arises from the
boundary condition $\frac {dT} {ds} = 0 $ at the same place. In
such structures the fundamental physical parameters pressure,
loop top temperature, and the loop length $L$ are not independent but are
related through so-called scaling laws viz.
\begin{equation}
p \ L \sim \ T_{max}^{\frac {11 - 2 \gamma} {4}} . \label{pm4}
\end{equation}
In order to make progress we assume {\it ad hoc}
that the distribution of the actual loop top temperatures $\Phi (T_{max})$
of a collection of loops can be described through a Gaussian viz.
\begin{equation}
\Phi (T_{max}) = \frac {1} {\sqrt{2 \pi} \sigma _T}
e^{- \frac {(T_{max} - T_0)^2} {2 \sigma _T ^2}},
 \label{pm5}
\end{equation}
with a mean loop top temperature $T_0$ and some dispersion $\sigma _T$.
We further assume - for the time being - that all loops have the same length.
It is clear that only loops with loop top temperature $T_{max} > T$
can contribute to the overall differential emission measure
at some given temperature $T$. The total differential emission
measure from such a collection of loops is therefore given by
\begin{eqnarray}
\xi_{tot} (T) \sim \frac {1} {L} T^{\alpha -\gamma/2 - 1/4}
\ \ \ \ \ \ \ \ \ \ \ \ \ \ \ \
\nonumber\\
\int _T ^{\infty} dT_{max}
T_{max}^{\frac {11 - 2 \gamma} {4} }
\frac {e^{- \frac {(T_{max} - T_0)^2} {2 \sigma _T ^2}}}
{\sqrt{1-(\frac {T} {T_{max}})^{2-\gamma + \beta}}}\,,
\label{pm6}
\end{eqnarray}
where we have used the scaling law Eq.~\ref{pm4} to replace the (constant)
loop pressure $p$ with $T_{max}$ and $L$. Because of the scaling law
Eq.~\ref{pm4} $p$ and $T_{max}$ are not independent, and in fact for given $L$
the pressure and thus the total emission measure depend quite
sensitively on $T_{max}$. If we then - finally - assume that the
loop length $L$ is related to the loop top temperature with some power
law coefficient $\delta$, we are led to the following analytical form of the
differential emission measure distribution:
\begin{eqnarray}
\xi_{tot} (T) = A_{norm} T^{\alpha _1}
 \ \ \ \ \ \ \ \  \ \ \ \ \ \ \ \  \ \ \ \ \ \ \ \  \ \ \ \ \ \ \ \
\nonumber\\
\int _T ^{\infty} dT_{max}
\frac {T_{max}^{\alpha _2}}
{\sqrt{1-(\frac {T} {T_{max}})^{\alpha _3}}}
e^{- \frac {(T_{max} - T_0)^2} {2 \sigma _T ^2}}\,,
\label{pm7}
\end{eqnarray}
with the three power law coefficients $\alpha_1$, $\alpha_2$, and
$\alpha_3$, the mean loop top temperature $T_0$ and its variance
$\sigma_T$ as parameters describing the shape of the distribution function.\\
With the {\it Chandra} LETGS data we can test whether the shape specified
by Eq.~\ref{pm7} is consistent with our observations. Calculations showed
that the hydrogen- to helium-like line ratios depend sensitively on
$\alpha_1$, $T_0$, and $\sigma _T$, but less sensitively on the parameters
$\alpha_2$ and $\alpha_3$. The best fit model was found to be
given by the set $\alpha_1$ = 0.48, $\alpha_2$ = 0.50, $\alpha_3$ = 0.50,
$T_0$ = 7\,MK, and $\sigma _T$ = 11\,MK. In Fig.~\ref{comp} we plot the DEM
distribution function in comparison to the best-fit DEM distributions
reconstructed from fourth-order Chebyshev polynomials with $T_{max}$ =
30\,MK and $T_{max}$ = 40\,MK; as is clear from Fig.~\ref{comp}, the DEM
reconstructions agree extremely well with each other for temperatures above
1\,MK, while the slope of the emission reconstructions disagrees for lower
temperatures. Since there are almost no constraints on the DEM in this
temperature range (note that the peak formation temperature for N\,{\sc vii}
is at 1\,MK), such a discrepancy is hardly surprising. As a first conclusion
we therefore note, that a function form as given in Eq.~\ref{pm7} provides a
reasonable description of the observed line ratios with suitably chosen
parameters $T_0$, $\sigma_T$, $\alpha_1$, $\alpha_2$, and $\alpha_3$. In the
framework of the chosen model it is a little surprising that the mean
(Gaussian) loop top temperature is rather low ($T=7$\,MK). However, because of
the large dispersion of 11\,MK, significantly hotter loops occur with high
probability. Since according to the scaling laws higher loop top temperatures
imply  higher pressure, higher pressure implies higher density, and higher
density implies larger emission measure, so the hotter loop will contribute
most. For constant cross section loops with the canonical radiative loss
function one expects $\alpha_1 \sim 0$; the observed value $\alpha_1
\sim 0.5 $ can only be explained by expanding loop geometries.

\section{Discussion and Conclusions}

The new generation of X-ray spectrometers on board the
{\it Chandra} and XMM-Newton
satellites allows the determination of elemental abundances in hot X-ray
emitting plasmas. The {\it Chandra} LETGS has the specific advantage of a
very large band pass with an ensuing sensitivity to lines from rather different
temperature regimes. Our analysis of the Algol LETGS spectrum shows
that the abundances for the elements neon, magnesium, silicon, and iron are all
sub-solar. This is in line with previously published high-resolution
abundance determinations of HR1099 with the {\it Chandra} HETGS
(Drake et al. \cite{drake01}) and XMM-Newton (Brinkmann et al.
\cite{bri2001}). Among those elements neon has the relatively highest
abundance, i.e., it is least sub-cosmic. It appears that these conclusions
are rather robust, and specifically do not sensitively depend on the methodology
used (``global fit'' vs. ``emission line analysis''). However,
an inspection of Tab.~\ref{abun} shows that, for example, the iron abundance
determinations considerably depend on the
lines used for the analysis. The short wavelength lines of Fe\,{\sc xvii}
at 15.01\,\AA, 15.27\,\AA, and 17.07\,\AA\ yield higher abundances than
the Fe lines at 93.92\,\AA, 108.37\,\AA, 128.37\,\AA, and
132.82\,\AA; the Fe line at 101.55\,\AA \ yields larger abundances than the 
rest of the EUV Fe lines.
The reason(s) for this discrepancy are not
quite clear. Optical depth effects in the Fe\,{\sc xvii} are very likely
not the cause (Ness et al. \cite{ness_opt}) and would
in fact even worsen the discrepancy. Since the short wavelength lines are all
from Fe\,{\sc xvii}, the calculated emission measure distribution for this
ion might be incorrect, i.e., too large. Since Fe\,{\sc xvii} is produced
over a rather large temperature range this appears somewhat unlikely
since then also the continuum emission would have to be incorrectly placed.
The Fe\,{\sc xvii} long wavelength lines are affected by absorption, but we
used already a rather large value of $N_H$; lowering the absorption
column density would again worsen the discrepancy. Systematic
errors in the instrument calibration might affect the long wavelength
portion of the spectrum in a different way
than the short wavelength portion, but
the magnitude of the effect is much larger than the systematic
calibration uncertainties ($\approx$ 15 \%). Finally, atomic physics
uncertainties might affect the long wavelength lines different compared to
the short wavelength lines. At any rate, we have to conclude and state
that we presently have no satisfactory explanation for the discrepant
abundance determinations for iron.

As to the abundance discrepancies derived from different analysis methods,
our studies have clearly demonstrated the importance of
the correct determination of the underlying temperature structure for a
correct determination of elemental abundances. The cooling functions of
individual lines contribute significantly over a temperature range of 0.3~dex
and the shape of the emission measure distribution also implies considerable
contributions far away from the peak formation temperatures of individual lines.
The lines used in our study, i.e., Ly$_{\alpha}$ and He-like r lines, are
formed over a rather wide temperature range, other lines, in particular lines
from ions with incompletely filled shells, are formed over somewhat narrower
temperature ranges necessitating an even better knowledge of the temperature
structure. We purposely used only those lines for our differential
emission measure reconstruction, because, first, these lines are among the
strongest for each element and therefore the most likely lines to be detected
in a recorded X-ray spectrum, and
second, the atomic physics of those lines ought to be known best. A reliable method
for abundance determination must prevent any cross talk between the temperature and
abundance structure of a plasma; therefore, the temperature structure should be determined
independent from the elemental abundances
either from line ratios of lines of the same chemical element (as done in this
paper) or by using lines only from the same element (as
done by, e.g., Drake et al. \cite{drake01}). We next 
emphasize the need of physical
considerations in the determination of the temperature structure. This is in
particular required if one ever desires to determine elemental abundances in
the X-ray range with an accuracy achieved by optical abundance determinations.
We have at the moment only few clues as to the differential emission measure
distributions realized by stellar coronae and
the uncertainties in our knowledge of the correct temperature
structure prevents from reaching precisely this goal.
A modeling of the coronal emission in terms of
individual temperature component is unsatisfactory from a physical point of
view, from a procedural point of view and from a mathematical point of
view. Abundances determined in this way may have small statistical errors
(of a few percent depending on the SNR of the modeled data), but rather
large systematic errors of 100 \% or more; nevertheless they are adequate to reveal
general trends in abundance patterns. This is exemplified in
Tab.~\ref{abunconv}, which shows that for example the oxygen abundance changes
by a factor two for models with discrete temperature components. A comparison
of the abundances derived for Algol in this paper with those derived by Antunes
et al. (\cite{ant94}) from ASCA using a two-temperature variable abundance
modeling approach also shows that the general trend in the run of elemental
abundances is captured and the "low" iron abundance and "high" neon abundance
are recognized, while the abundances of individual elements can vary by at least a
factor of two. Also, the real clue of the {\it Chandra} LETGS Algol
observation, i.e., the overabundance of nitrogen with its profound physical
implications (cf. Schmitt and Ness \cite{schm2002}) went unnoticed in the
modeling with the lower resolution ASCA data.

As to Algol specifically, our detailed temperature and abundance modeling
confirms the results previously derived by Schmitt and Ness (\cite{schm2002}).
Because of temperature dependence of the emissivity functions of the
Ly$_{\alpha}$-lines for C and N (cf. Fig.~\ref{em_theo}), the line
ratio between these lines must stay below 0.57 (for cosmic abundances)
regardless of the underlying temperature structure in contrast
to the observed ratio of $>$ 23. Our modeling now shows that carbon
is depleted down to at least 8\,\%, while nitrogen is enhanced by about
70\% or more (all relative to cosmic abundances). This effect is dramatic.
Assuming the cosmic abundance pattern recommended by Holweger
(\cite {hol01}) there are 4.58 carbon atoms for every nitrogen atom,
while in Algol's corona we have (at least) 8.2 nitrogen atoms for
every carbon atom ! This reversal of carbon and nitrogen abundance
can be readily explained by assuming that one is studying CNO-cycle
processed material in the corona of Algol B, since the equilibrium abundance
of CNO nuclei participating in the cycle is such that most nuclei occur
as $N^{14}$ nuclei (Caughlan \cite{cau65}).
In no other spectral range than the X-ray band
can the chemical abundance of the B component of the Algol system be studied.

Our {\it Chandra} LETGS spectrum of Algol thus demonstrates the wealth of
physical information contained in an X-ray spectrum with high spectral resolution
and - at the same time - good signal-to-noise ratio. The latter is as important
as the former, since data with poor signal-to-noise will not allow the
derivation of meaningful and significant results. The exposure of such spectra
requires substantial satellite resources, yet it represents the only way to
extract information on the physics of stellar coronae.

\begin{table*}
\caption[ ]{\label{sum1tab} Comparison of coronal abundances for Algol derived
from {\it Chandra} LETGS (this paper, second column) with abundances derived
from ASCA (third to fifth column; Antunes et al. \cite{ant94}) and EUVE
(sixth column; Stern et al. \cite{stern95}).}

\begin{tabular}{|c|r|c|c|c|c|}
\hline
Element & This paper & Antunes et al. & Antunes et al. & Antunes et al. & Stern et al.\cr
        &            & Low State & Medium State & High State & \cr
\hline
C  & $<$ 0.04 &  n.a. & n.a. &  n.a. & n.a. \cr
N  & 2.0  &  $<$ 0.1  & $<$ 0.1  & $<$ 0.1  & n.a. \cr
O  & 0.25  & 0.30 $\pm$ 0.04 & 0.31 $\pm$ 0.03 &  0.24 $\pm$ 0.03 & n.a. \cr
Ne & 0.95  & 0.76 $\pm$ 0.10 & 1.22 $\pm$ 0.08 & 1.08 $\pm$ 0.08  & n.a. \cr
Mg & 0.5  & 0.48 $\pm$ 0.06 & 0.64 $\pm$ 0.05 & 0.47 $\pm$ 0.04 &n.a. \cr
Si & 0.45  & 0.43 $\pm$ 0.05 & 0.65 $\pm$ 0.04 & 0.47 $\pm$ 0.03 & n.a. \cr
Fe & 0.2  & 0.30 $\pm$ 0.01 & 0.37 $\pm$ 0.02 & 0.32 $\pm$ 0.01& 0.2-0.4\cr
\hline
\end{tabular}

\end{table*}

\begin{acknowledgements}
J.-U. Ness acknowledges suppert from the DLR grant 50OR0105. We acknowledge useful discussions
and help from Drs. P. Predehl and V. Burwitz.
\end{acknowledgements}

\end{document}